\documentclass[11pt,twoside]{article}

\usepackage{asp2006}
\usepackage{epsf}
\usepackage{lscape}
\usepackage{natbib}

\begin{document}
\title{Morphologies of the Nebulae around ``born-again'' Central Stars of Planetary Nebulae}   %%% Fill in title

\author{S. Kimeswenger,$^1$
A.A. Zijlstra,$^2$ P.A.M. van Hoof,$^3$ M. Hajduk,$^4$ \\F.
Herwig,$^5$ M.F.M. Lechner,$^1$ S.P.S. Eyres,$^6$ and G.C. Van de
Steene$^3$}

\affil{$^1$Institute of Astro and Particle Physics, University
Innsbruck, \protect\newline\phantom{i}Technikerstra{\ss}e 25, 6020
Innsbruck, Austria}

\affil{$^2$University of Manchester, School of Physics \&\
Astronomy, \protect\newline\phantom{i}P.O. Box 88, Manchester M60
1QD, UK}

\affil{$^3$Royal Observatory of Belgium, Ringlaan 3, 1180 Brussels,
Belgium}

\affil{$^4$Centrum Astronomii UMK, ul.Gagarina 11, PL-87-100 Torun,
Poland}

\affil{$^5$Astrophysics Group, School of Physical and Geographical
Sciences, \protect\newline\phantom{i}Keele University, Staffordshire
ST5 5BG, UK}

\affil{$^6$Centre for Astrophysics, University of Central
Lancashire, Preston \protect\newline\phantom{i}PR1 2HE, UK}

\begin{abstract} %%% Abstract to run on from here.
While in the past spherodicity was assumed, and still is used in
modeling of most nebulae, we know now that only a small number of
planetary nebulae (PNe) are really spherical or at least nearly
round. Round planetary nebulae are the minority of objects. In the
case of those objects that underwent a very late helium flash
(called VLTP objects or ``born-again'' PNe) it seems to be
different. The first, hydrogen-rich PN, is more or less round. The
ejecta from the VLTP event, in contrast, are extremely asymmetrical.
%%Angular momentum is mostly
%%assumed to be the main reason for the asymmetry in PNe. Thus we have
%%to find processes either changing their behavior within a few
%%hundred to a few thousands of years or change their properties
%%dramatically due to the variation of the abundance.
\end{abstract}

%%% MAIN BODY OF TEXT GOES HERE. CONSULT "INSTRUCTIONS FOR AUTHORS USING
%%% LATEX2E MARKUP", SECTIONS 2.3-2.6 FOR HELP WITH EQUATIONS, FIGURES,
%%% AND TABLES.

\section{The VLTP nebulae}
The known family of VLTP objects is rather small. Only a few show a
clear signature of a VLTP event \citep[first described by][]{vltp1}: a
hydrogen rich normal PN and prominent hydrogen poor ejecta near the
core - and if observable - a hydrogen poor central star
(CSPN). Prominent members from the review by \citet{albert} are
presented here to point out their common morphologies. Below, we show
a possible way to search for similar objects by means of their global
properties.

\subsection{A30 \& A78: the Seniors}

\begin{figure}[!ht]
%%\centering
\plotfiddle{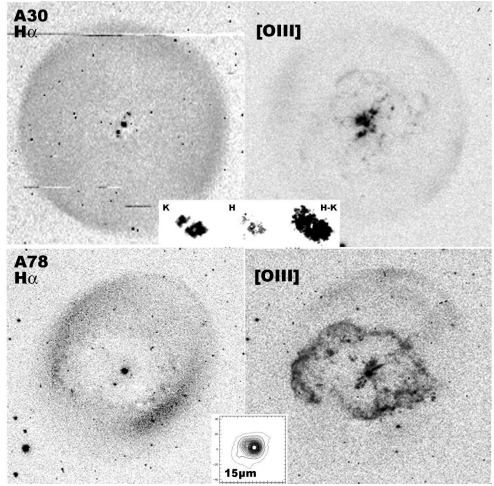}{5.5cm}{0}{34}{34}{-90}{-10}
\caption{Abell 30 (top) and Abell 78 (bottom): The H$\alpha$ image
(A30 by \cite{balick}, A78 from Calar Alto) and the [OIII] images
(ING archive and Calar Alto) show the perfectly round shape of the
hydrogen-rich old PN and the clumpy ejecta. The inserts show
infrared data: K \& H from \cite{ks_97_a30}, ISO 15$\mu$m from
\cite{ks_98}. They indicate a belt of hot dusty material
perpendicular to the main axis as defined by the VLTP ejecta in
[OIII].} \label{kimeswenger:fig:1}
\end{figure}

A30 (GPN G208.55+33.28) is --probably-- the oldest, and best studied
member of the family.  \cite{Jacoby_Ford83} discovered the unusual
abundance of the central knots. A30 shows all features of a
``born-again'' PN (see Fig.~\ref{kimeswenger:fig:1}) and can be
considered as the prototype of this class:
\begin{itemize}
\item A hydrogen poor central star with prominent C and O wind emission
lines and unusual wind properties
%%(\citeauthor{kaler84}\citeyear{kaler84},
%%\citeauthor{wind1}\citeyear{wind1}).
\citep{wind1}.
\item \vspace{-3mm}Fast moving hydrogen-poor knots embedded in a normal old PN
%%(\citeauthor{Jacoby_Ford83}\citeyear{Jacoby_Ford83},
%%\citeauthor{reay83}\citeyear{reay83}).
\citep{Jacoby_Ford83}.
\item \vspace{-3mm}Unusual hot and small carbonous dust grains near the core
in a
ring/belt-like structure
%%(\citeauthor{dust1}\citeyear{dust1},
%\citeauthor{dust2}\citeyear{dust2})
\citep{dust1}.
\end{itemize}

The old nebula is a perfect --and rare--example of a round PN. The
hydrogen-poor and dusty ejecta are, in contrast, strongly
bipolar(Fig.~\ref{kimeswenger:fig:1}, inserts)

The old nebula of A78 (GPN G081.29-14.91) also shows considerable
symmetry. It appears more barrel-shaped, inclined to the line of
sight. But this view is overemphasized by an excitation effect.
[OIII] is more prominent at the poles. In contrast to the symmetry
of the outer nebula, \cite{ks_98} show in their ISO study a belt of
dusty material perpendicular to the main direction of the fast
ejecta. In this case the clumpy ring extends into the area of the
old PN. Thus, in both cases the old nebula is far more spherical
than the newly ejected VLTP material.

\subsection{V605 Aql \& V4334 Sgr: Bipolarity at young ages}
\begin{figure}[!ht]
%%\centering
\plotfiddle{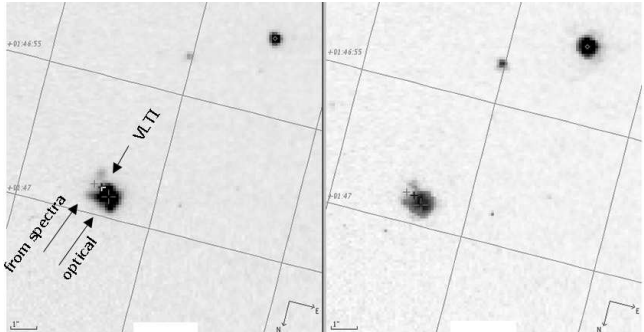}{3.5cm}{0}{60}{60}{-135}{-10}
%%\plotone{fig2tue_b.eps}
 \caption{The HST [NII] and [OIII] images of A58
with a re-calibrated coordinate system. The center derived by
spectra in \citet{LaPalma07} and by one of us (MH) using VLA data is
clearly offset to the optical center of the knots.}
\label{kimeswenger:fig:2}
\end{figure}

V4334 Sgr (Sakurai's object; the CSPN of GPN G010.47+04.41) and V605
Aql (the CSPN of A58 = GPN G037.60$-$05.16) underwent VLTP events
within the past century. Their outer nebulae show a large degree of
spherical symmetry, for A58 affected by ISM interaction
\citep{pnism}.  The inner cores show spectroscopic evidence for
bipolarity, as shown in more detail in \citet{LaPalma07}. The core of
A58 also shows position offsets of the central emission at different
wavelengths (Fig.~\ref{kimeswenger:fig:2}), indicative of an
extinction torus.  For V605 Aql such a bipolarity of the dust
distribution already has been suspected earlier by
\citet{Pollacco1992} and \citet{nilfisc}. The asymmetric shell of
V4334 Sgr became visible during the second and third year after the
return to the cold luminous state: it was monitored in detail by
\citet{Kimeswenger DUST}.
% The HST images of V605 Aql, re-calibrated with respect to
% their coordinate system using UCAC stars are shown in
% Fig.~\ref{kimeswenger:fig:2}.

\section{Discussion}

The strong bipolarity of the VLTP events requires a common
explanation.  A binary-induced shaping, or stellar rotation, is
less likely, first because of the nature of the VLTP event, and second
because of the absence of strong shaping in the old ejecta.  The
bipolarity can be amplified during the transition from the carbon-rich
post-AGB like state to the early hot core when a fast hot wind crashes
into the high extinction VLTP ejecta \citep[see e.g.][]{Icke}. This
requires an initial asymmetry during the VLTP flash.  Fully convective
configurations are dominated by a dipole convective flow structure
\citep{convection1,convection2}. One may speculate that the expanding
VLTP convection may equally be dominated by a dipolar flow, since the
convectively stable core in the middle will eventually occupy only a
small fraction of the expanding star.

%\section{The Search}

\begin{figure}[!ht]
%%\centering
\plotfiddle{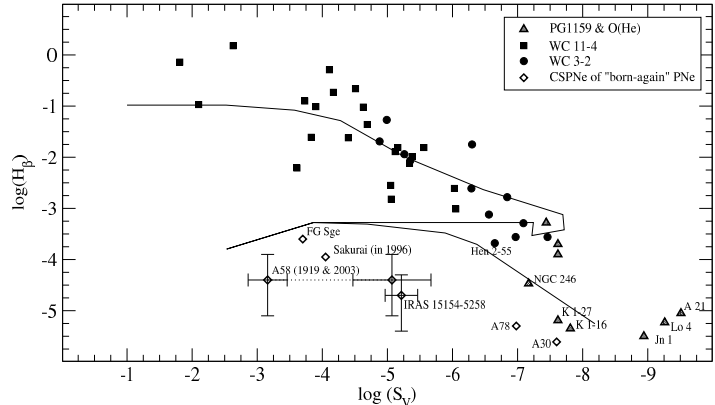}{4.5cm}{0}{40}{40}{-150}{-30}
%%\plotone{fig3tue.eps}
\caption{The diagram of the nebular surface brightness $S_{\rm
H\beta}$ (corresponding to the evolution of the old nebula) vs. the
surface brightness calculated from the CSPN flux $S_{\rm V}$
(corresponding to the evolution of the CSPN). For a detailed
description of the graph see \citet{GoTy00}.}
\label{kimeswenger:fig:3}
\end{figure}

The family of old VLTP members is probably incomplete.
\citet{GoTy00} list 65 hydrogen poor central stars of PNe. Hot
PG1159 white dwarfs are also thought to be remnants of this kind of
event. Less than half of them have a PN, but most of those having a
PN do currently not show (prominent) hydrogen poor ejecta. \\
Beside detailed morphological analysis, we propose the use of the
average nebular surface brightness $S_{\rm H\beta}$ vs. the surface
brightness calculated from the stellar flux over the apparent size
of the nebula $S_{\rm V}$ \citep[see][]{GoTy00}, to locate more
``born-again'' nebulae (Fig.~\ref{kimeswenger:fig:3}). For V605
Aql/A58 the flux of the current CSPN was calculated from predicted
values of \citet{herwig}. For objects lacking an $H_\beta$ flux we
estimated it from the radio flux.  While the old nebula dominating
the vertical axis of this graph passes through its undisturbed
evolution, the CSPN undergoes a rapid evolution. Thus the VLTP
objects can be found far from the regular PNe \citep[see][for the
whole sample of normal PNe in this kind of graph]{GoTy00}.\\ The
known ``born-again'' nebulae occupy a region below the track given
by the original authors based on the track by \citet{vltp83}.  They
are well separated from those PNe having a PG1159 core without
having hydrogen-poor dusty ejecta in the nebula. Thus may we have to
think about two different branches of evolution -- with and without
a dusty phase ?

%\section{Summary}
%It seems to be a common morphology, that the born again ejecta are
%placed along a strong bipolar axis independent of the appearance of
%the old (first) nebula. This might have its origin in colliding
%streams of material reflected from the inner edges of a dusty torus
%\citep{Icke}. Such a torus might have been built during the strong
%dust formation phase of the carbon rich ``born-again'' cold giant
%star near the AGB in the HR diagram.

\vspace{-1mm}
\acknowledgements %%% Text of acknowledgements runs on after this command.
PvH is supported by the Belgian Science Policy grant MO/33/017.


\vspace{-1mm}

\begin{thebibliography}{}
\bibitem[Balick(1987)]{balick} Balick, B. 1987, AJ 94, 67
\bibitem[Icke(2007)]{Icke}
Icke, V. 2007, in Asymetric Planetary Nebulae IV, ed. R.L.M.
Corradi, A. Manchado and N. Soker. (in press)
\bibitem[Borkowski et al.(1994)]{dust1}Borkowski, K.J., Harrington, J.P., Blair, W.P., \& Bregman, J.D. 1994, ApJ, 435, 722
%%\bibitem[Colangeli et al.(1993)]{dust2}Colangeli, L., Mennella, V., Blanco, A., et al. 1993, ApJ, 418, 435
%%\bibitem[Corsico et al. (2007)]{werner07} Corsico, A.H., Althaus, L.G., Miller-Bertolami, M.M., \& Werner, K. 2007, A\&A, 461, 1095
%\bibitem{speck}C. Dijkstra, A.K. Speck: ApJ \textbf{651}, 288
%(2006)
\bibitem[G{\'o}rny \& Tylenda(2000)]{GoTy00}G{\'o}rny, S.K., \& Tylenda, R. 2000, A\&A, 362, 1008
%\bibitem{Hajduk2007}M. Hajduk, A.A. Zijlstra, P.A.M. van Hoof, et
%al: MNRAS \textbf{378}, 1298 (2007)
%\bibitem{Science}M. Hajduk, A.A. Zijlstra, F. Herwig, et al: Science \textbf{308}, 231 (2005)
\bibitem[Herwig(2001)]{herwig}Herwig, F. 2001, ApJ, 554, L71
%\bibitem[Herwig (2007)]{herwig07} Herwig, F. 2007, in Hydrogen Deficient Stars, ed. K. Werner and T.
%Rauch, ASP Conf. Ser. (this volume)
\bibitem[Iben et al.(1983)]{vltp83}Iben Jr., I., Kaler, J.B., Truran, J.W., \& Renzini, A. 1983, ApJ, 264, 605
%%\bibitem[Jacoby \& Ford (1981)]{Jacoby_Ford81} Jacoby, G.H. \& Ford H.C. 1981, BAAS, 13, 854
\bibitem[Jacoby \& Ford(1983)]{Jacoby_Ford83}Jacoby, G.H.,  \& Ford, H.C. 1983, ApJ, 554, L71
%%\bibitem[Kaler \& Feibelman(1984)]{kaler84} Kaler, J.B., \& Feibelman, W.A. 1984, ApJ, 282, 719
\bibitem[Kimeswenger \& Koller(2002)]{Kimeswenger DUST}Kimeswenger, S., Koller, J. 2002, ApSS,
279, 149

%\bibitem{ks2001} S. Kimeswenger: Rev. Mex. Astron. \& Astrop. \textbf{37},
%115 (2001)
\bibitem[Kimeswenger et al.(1997)]{ks_97_a30} Kimeswenger, S., Kerber, F., \& Roth, M. 1997, AG Abstr. Ser., 13,
227
\bibitem[Kimeswenger et al.(1998)]{ks_98} Kimeswenger, S., Kerber, F., \& Weinberger, R. 1998, MNRAS, 296,
614
\bibitem[Kimeswenger et al.(2007)]{LaPalma07}
Kimeswenger, S., Zijlstra, A.A., van Hoof, P.A.M., Hajduk, M.,
 Lechner, M.F.M., Van de Steene, G.C., \& Gesicki, K.
 2007, in Asymetric Planetary Nebulae IV, ed. R.L.M.
Corradi, A. Manchado and N. Soker. (in press)
\bibitem[Koller \& Kimeswenger(2001)]{nilfisc} Koller, J., \& Kimeswenger, S. 2001, ApJ, 559,
419
%\bibitem{manchado89} A. Manchado, P. Garc{\'i}a-Lario, S.R. Pottasch: A\&A \textbf{218}, 267 (1989)
%\bibitem{naylor}
%T. Naylor, P.A. Charles, K. Mukai, A. Evans: MNRAS \textbf{258}, 449
%(1992)
\bibitem[Kuhlen et al.(2006)]{convection1}Kuhlen, M., Woosley, S.E., \& {Glatzmaier}, G.A. 2006, ApJ, 640, 407
\bibitem[Pollacco et al.(1992)]{Pollacco1992} Pollaco, D.L.,  Lawson, W.A.,  Clegg, R.E.S.,
\& Hill P.W. 1992, MNRAS, 257, p33
\bibitem[Porter \& Woodward(2000)]{convection2}{{Porter}, D.H., \& {Woodward}, P.R.} 2000, ApJS, 127, 159
%%\bibitem[Reay et al.(1983)]{reay83}Reay, N.K.,  Atherton, P.D., \& Taylor, K. 1983, MNRAS, 203, 1079
%\bibitem{Seitter}W.C. Seiter: Mitt. Astron. Ges. \textbf{63}, 181 (1985)
%\bibitem{SakLetter}P.A.M. van Hoof, M. Hajduk, A.A. Zijlstra, et al: A\&A \textbf{471}, L9 (2007)
%\bibitem{WoFa73}P.R. Wood, D.J. Faulkner: ApJ \textbf{181}, 147
%(1973)
\bibitem[Sch{\"o}nberner(1979)]{vltp1} Sch{\"o}nberner, D. 1979,
A\&A, 79, 108
\bibitem[Wareing et al.(2007)]{pnism} Wareing, C.~J.,
Zijlstra, A.~A., \& O'Brien, T.~J.\ 2007, MNRAS, in press (arXiv0709.2848)
arXiv:0709.2848 \bibitem[Yadoumaru \& Tamura(1994)]{wind1} Yadoumaru, Y.,  \& Tamura, S. 1994, PASP, 106, 165
\bibitem[Zijlstra(2002)]{albert}Zijlstra, A.A. 2002, Ap\&SS, 279, 171

\end{thebibliography}
\end{document}